\documentclass[%
 showkeys,
 reprint,
 amsmath,amssymb,
 aps,
]{revtex4-2}

\usepackage{graphicx}% Include figure files
\usepackage{dcolumn}% Align table columns on decimal point
\usepackage{bm}% bold math
\begin{document}

\preprint{}

\title{Monte Carlo renormalization group calculation for the d=3 Ising Model using a modified transformation}

\author{Dorit Ron}
\altaffiliation{dorit.ron@weizmann.ac.il}
\author{Achi Brandt}%
\email{achi.brandt@weizmann.ac.il}
\affiliation{%
 Faculty of Mathematics and Computer Science, The Weizmann Institute of Science, Rehovot 76100, Israel
}%

\author{Robert H. Swendsen}
\homepage{swendsen@cmu.edu}
\affiliation{
 Department of Physics, Carnegie Mellon University, Pittsburgh, Pennsylvania, 15213, USA
}%

\date{\today}

\begin{abstract}
We present a simple approach to high-accuracy calculations of critical properties for
the three-dimensional Ising model, without prior knowledge of the critical temperature.
The iterative method uses a modified block-spin transformation with a tunable parameter
to improve convergence in the Monte Carlo renormalization group trajectory.
We found experimentally that the iterative method enables the calculation of the critical temperature simultaneously with a critical exponent. 

\end{abstract}

\keywords{Critical exponents; Monte Carlo renormalization group; optimized convergence}

\maketitle

\section{Introduction}\label{section: Introduction}

The Monte Carlo renormalization group (MCRG) method
is a systematic procedure for computing critical properties
of lattice spin models\cite{Ma_MCRG_1976,RHS_MCRG_1979}.
It has been shown to be both flexible and effective
in the calculation of critical exponents, critical temperatures,
and renormalized couplings constants\cite{RHS_MCRG_optimize_d=2,RHS_MCRG_optimize_d=3,Brandt_Ron_2001,Ron_RHS_2001,Ron_RHS_2002,Ron_RHS_Brandt_2002,Ron_RHS_Brandt_2005}.

A particularly interesting application of MCRG is the three-dimensional
Ising model\cite{Bloete_RHS_3d_Ising_1979,Pawley_MCRG_3d_Ising_1984}.
This model has proven to be one of the most difficult to obtain accurate results for,
because the approach to the fixed point is so slow.

The most encouraging result has been that of
Bl\"ote et al.\cite{Bloete_Heringa_Hoogland_Meyer_Smit_1996}
who used a three-parameter approximation to the fixed point,
along with a modified majority rule for the RG transformation.
The parameter involved in this modified rule can be further tuned
to calculate the critical exponents by accelerating the convergence of
the RG transformation as was demonstrated  in\cite{Ron_Brandt_RHS_2017}.
There, we determined the optimal value of the parameter for the calculation of the largest even and odd eigenvalues, $y_{T1}$ and $y_{H1}$, much more carefully and accurately than in earlier work and showed its fast convergence compared with that of the majority rule.

In this work, we show that this accelerated convergence can be further exploited to also directly determine the inverse critical temperature, which is usually unknown for a general model. We present a simple iterative method which enables the simultaneous calculation of the inverse critical temperature along with 
the largest odd eigenvalue exponent $y_{H1}$.

In the following section we recall the MCRG method.   In section \ref{tunable RG} we review the tunable RG transformation, while in section \ref{other parameterization} we introduce another possible parameterization for the RG transformation. 
In section \ref{Tc and yH1} we present our approach along with the results for the simultaneous calculations of the inverse critical temperature and for $y_{H1}$.
Finally, we present our conclusions and discuss future work.

\section{MCRG computations}

The MCRG has often been reviewed\cite{Ma_MCRG_1976,RHS_MCRG_1979,RHS_MCRG_optimize_d=3,Baillie_Gupta_Hawick_Pawley_1992}, and here we only briefly outline the method. 
We consider the three-dimensional Ising model on a simple periodic cubic lattice,
of size $N \times N \times N$.
The Hamiltonian is given by
\begin{equation}\label{NN H}
H = K \sum_{\langle j,k \rangle} \sigma_j    \sigma_k    ,
\end{equation}
where
$\sigma_j = \pm 1$,
and
the sum is over all nearest-neighbor pairs.
The dimensionless coupling constant
$K$ has been defined to include the inverse temperature
$\beta=1/k_B T$,
so as to make the Boltzmann factor $e^H$.

The model was simulated using the Wolff algorithm\cite{Wolff_1989}
to generate a set of configurations
characterizing the equilibrium distribution.

The renormalized configurations are obtained from these sets.
For each configuration, the lattice is divided up into   cubes, each containing eight sites,
so that the scaling factor $b=2$.
A value of plus or minus one is assigned to each renormalized spin to represent the  original spins in each cube in a way described below.

It is convenient to write the starting Hamiltonian (original simulated system) Eq.~(\ref{NN H}) in its most general form:

\begin{equation}\label{General H}
H^{(n)} =  \sum_{\alpha} K_{\alpha}^{(n)} S_{\alpha}^{(n)}        ,
\end{equation}
where the interactions $S$'s are combinations of the spins and the $K$'s are the corresponding coupling constants. The sum is over all possible interactions that exist on a lattice of a given size.
The subscript $\alpha$ denotes the type of interaction or coupling
(nearest-neighbor, next-nearest-neighbor, four-spin, etc.). The superscript ${n}$ is the number of applied renormalization steps.
Since we have just described the first iteration of the renormalization
transformation, $n=1$.
The nearest-neighbor coupling constant $K$ defined earlier in Eq.~(\ref{NN H}),
will also be denoted by
$K_{nn}^{(0)}$.
All other coupling constants at level $n=0$  vanish.

To determine the critical exponents,
we then need to calculate
the matrix of derivatives of the couplings at level $n+1$
with respect to the couplings at level $n$.
\begin{equation}\label{T-matrix}
T_{\alpha,\beta}^{(n+1,n)}
=
\frac{
\partial  K_{\alpha}^{(n+1)}
}{
\partial  K_{\beta}^{(n)}  
}   ~.
\end{equation}
This matrix of derivatives is
 then given by the solution of  the equation
\begin{equation}
\frac{ \partial \langle  S_{\gamma}^{(n+1)} \rangle  }
{\partial  K_{\beta}^{(n)} }
=
\sum_{\alpha}
\frac{ \partial \langle  S_{\gamma}^{(n+1)} \rangle  }
{\partial  K_{\alpha}^{(n+1)} }
\frac
{\partial  K_{\alpha}^{(n+1)} }
{\partial  K_{\beta}^{(n)} }   ~,
\end{equation}
where
\begin{equation}
\frac{ \partial \langle  S_{\gamma}^{(n+1)} \rangle  }
{\partial  K_{\beta}^{(n)} }
=
\left\langle  S_{\gamma}^{(n+1)}  S_{\beta}^{(n)}  \right\rangle
-
\left\langle  S_{\gamma}^{(n+1)}  \right\rangle
\left\langle  S_{\beta}^{(n)}  \right\rangle   ,
\end{equation}
and
\begin{equation}
\frac{ \partial \langle  S_{\gamma}^{(n+1)} \rangle  }
{\partial  K_{\alpha}^{(n+1)} }
=
\left\langle  S_{\gamma}^{(n+1)}  S_{\alpha}^{(n+1)}  \right\rangle
-
\left\langle  S_{\gamma}^{(n+1)}   \right\rangle
\left\langle S_{\alpha}^{(n+1)}  \right\rangle   .
\end{equation}

For our calculations  we have included $N_o=20$ odd interactions.
We have followed\cite{Baillie_Gupta_Hawick_Pawley_1992},
who calculated all $53$ even and $46$ odd interactions that fit in either a $3 \times 3$ square or a $2 \times 2 \times 2$ cube of spins, and used their first $20$ odd interactions.
The eigenvalues of the $T$-matrix
in Eq.~(\ref{T-matrix})
are found separately for the even and odd interactions.
The largest odd eigenvalue exponent $y_{H1}$ calculated below is then obtained from the largest eigenvalue of the odd $T$-matrix by $y_{H1}={ln\lambda}/{ln2}$, as $b = 2$.

\section{Tunable block-spin transformation}\label{tunable RG}

In\cite{Ron_Brandt_RHS_2017}, we showed that the usual majority rule, which performs well for the two-dimensional Ising model, converges very slowly for the three-dimensional Ising model.

So, instead of using the usual majority rule, the renormalized spin $\sigma'_{\ell}= \pm 1$, associated with $\ell$, a $2 \times 2 \times 2$ cube of spins, was assigned a value according
to the following probability\cite{Bloete_Heringa_Hoogland_Meyer_Smit_1996}:

\begin{equation}\label{RG_transform}
P( \sigma'_{\ell} ) = \frac{
\exp ( w \, \sigma'_{\ell} \sum_{j \in \ell} \sigma_j  )
}{
\exp ( w \sum_{j \in \ell}  \sigma_j  )  +  \exp ( - w \sum_{j \in \ell}  \sigma_j )
}~.
\end{equation}
For $w \rightarrow \infty$, this tends to the majority rule.

A special feature of this calculation is that the convergence of the RG transformation
in Eq.~(\ref{RG_transform})
can be enhanced by optimizing the parameter $w$ separately for each exponent.
In\cite{Ron_Brandt_RHS_2017}, we determined the optimal value of $w$
for the calculation of $y_{T1}$ and $y_{H1}$ much more carefully than in earlier work and showed its fast convergence compared with that of the majority rule.
We found
$w(y_{T1})=0.4314$
and
$w(y_{H1})=0.555$.
The determination of $w$ for $y_{T2}$ and $y_{H2}$ turned out to need much more statistics.
We observed that using the $w$ obtained for $y_{T1}$ ($y_{H1}$) also for the calculation of $y_{T2}$ ($y_{H2}$),
looked promising, but we believe that accuracy can still be enhanced.

The value of $w$ was adjusted
so that the sequence of calculated exponents would converge as fast as possible. 
That is, one may aim at vanishing differences $d_{{n+1},{n}}=0$,
\begin{equation}\label{dnp1 n}
d_{{n+1},{n}} =  \mbox{exponent }^{(n+1)}   - \mbox{exponent }^{(n)} = 0 ,
\end{equation}
where $n$ denotes the number of RG iterations. 
Our final results showed little dependence of the exponent estimates
on the number of RG iterations,
and the very small fluctuations that remained did not appear to be systematic.
We decided that an attempt to further reduce the errors was not promising.

\section{Other parameterization}\label{other parameterization}
It is important to note that there is nothing special in the from of Eq.~(\ref{RG_transform}) and thus it is possible to choose other parameterizations. For instance, one can
parameterize $P( \sigma'_{\ell} )$ by

\begin{equation}\label{tunable-w}
P( \sigma'_{\ell} ) =  \left\{
  \begin{array}{llll}
  1 & \mbox{if } \sum_{j \in \ell}  \sigma_j=4,6,8  \\
  w' & \mbox{if } \sum_{j \in \ell}  \sigma_j=2 \\
  0.5 & \mbox{if } \sum_{j \in \ell}  \sigma_j=0 \\
  1 - P( \sigma'_{\ell} ) & \mbox{if } \sum_{j \in \ell}  \sigma_j<0 ~~~.
  \end{array} \right.
\end{equation}

Just to demonstrate this point, we did some tuning of the parameter  $w'$ for lattice size $128^3$ at  $K_c=0.2216544$\cite{Talapov_Blote_1996}. We got for the odd eigenvalue exponent $y_{H1}$ similar results to those obtained by using Eq.~(\ref{RG_transform}) as introduced in\cite{Ron_Brandt_RHS_2017}, e.g., the exponent we obtained for $N_o=20$ after two renormalizations using $w'=0.88$ was $2.4828(1)$, compared with $2.4829(2)$ there.

\section{Tuning the block-spin transformation along with the inverse critical temperature $K_c$}
\label{Tc and yH1}

In\cite{Ron_Brandt_RHS_2017}, we used a known approximation to the critical inverse temperature $K_c=0.2216544$  and showed that upon tuning the block-spin transformation parameter $w$, a faster convergence to the fixed point value of the critical exponent was achieved and hence better estimations for the critical exponent were obtained. Since for a general model the inverse temperature is usually unknown, we want to present a method in which both parameters, i.e., $K_c$ and the optimal $w$ can be simultaneously calculated, and thus enabling the fast extraction of the largest odd eigenvalue exponent $y_{H1}$. First, instead of using $w$ (Eq.~(\ref{RG_transform})), we use
\begin{equation}\label{nu}
u = \frac{
1
}{
(1 + exp ( -4 w ) )
}~,
\end{equation}
which behaves somewhat more linear than $w$ as $K_c$ is approached.
We assume  $y_{H1}$ is a function of both $K_c$ and $u$, i.e.,

\begin{equation}\label{interp6}
 y_{H1} = a u + b u^2 + c K_c + d u K_c + e u^2 K_c + f.
\end{equation}

We observed that while finding an optimal $w$, namely a $w$ for which $d_{{n},{n-1}}=0$  (see Eq.~(\ref{dnp1 n})), for the largest $n$ we used, the value of $d_{{n-1},{n-2}}$ at that $w$, was also very small. This was the case {\it only} when the known critical value of $K_c$ was used. We experimentally show below that the two conditions $d_{{n},{n-1}}=d_{{n-1},{n-2}}=0$ may jointly be used for determining $K_c$ and $u$ (and hence $w$), at least with very good accuracy. For large enough grids more such conditions may be minimized by using least square calculations as well as using a better approximation, such as,

\begin{equation}\label{interp7}
 y_{H1} = a u + b u^2 + c K_c + d u K_c + e u^2 K_c + f + g K^2_c.
\end{equation}

We applied this approach to lattices $128^3$ and $256^3$.
From the six first lines of Table \ref{table: YH1 K and w_Wolff L=7}
we may calculate the six coefficients appearing in Eq.~(\ref{interp6}) separately for each $n=2, 3, 4$ and $5$. Then by demanding $d_{3,2}=d_{4,3}=d_{5,4}=0$, a new pair of $K_c$ and $u$ (and hence $w$) are obtained by the Newton-Raphson method. See line 7 in Table \ref{table: YH1 K and w_Wolff L=7} for lattice size $128^3$. Two additional iterations are shown in line 8(9), where we used Eq.~(\ref{interp7}) for the first six lines together with line 7(8, respectively). The resulting approximation for $K_c$ is $K_c=0.2216541(1)$  
along with our estimation for $y_{H1}=2.4819(1)$.

Table \ref{table: YH1 K and w_Wolff L=8} shows similar results for lattice $256^3$, where we obtained
$K_c=0.2216547(1)$ and $y_{H1}=2.4824(1)$. 

\begin{table*}[htb]
%\small
\caption{The odd eigenvalue exponent $y_{H1}$ calculated on lattice $128^3$ by using the
Wolff algorithm\cite{Wolff_1989} with different values of $K$ and $u$. $n$ denotes the number of RG iterations.
}
\begin{center}
\begin{ruledtabular}
\begin{tabular}{cccccccc}
%\hline
$ $ & $K$  & $w$ & $u$ & $n=5$ & $n=4$ & $n=3$ & $n=2$ \\
\hline						
1 & 0.2216 & 0.5	& 0.8808 & 2.44676(12)  &  2.47269(7)   &   2.48294(3) &  2.49179(2) \\	
2 & 0.2216 & 0.6	& 0.9168 & 2.44690(14)  &  2.47009(6)	&   2.47648(3) &  2.47887(2)  \\		
3 & 0.2216 & 0.7	& 0.9427 & 2.44725(14)  &  2.46895(6)	&   2.47335(3) &  2.47158(2)  \\		
4 & 0.2217 & 0.5	& 0.8808 & 2.51531(13)  &  2.49572(6)	&   2.49065(2) &  2.49430(2)  \\
5 & 0.2217 & 0.6	& 0.9168 & 2.51190(15)  &  2.49178(5)	&   2.48378(2) &  2.48133(1)  \\		
6 & 0.2217 & 0.7	& 0.9427 & 2.51031(14)  &  2.48999(5)	&   2.48041(2) &  2.47397(2)  \\
\hline
7 & 0.22165322 & 0.58819 & 0.9132  & 2.48122(10)  &  2.48181(4)	&   2.48096(2) &  2.48131(1)  \\
8 & 0.221654013 & 0.58674 & 0.912692 & 2.48167(10)  &  2.48199(4)	&   2.48110(2) &  2.48147(1)  \\
9 & 0.22165417 & 0.58628 & 0.912546 & & & & \\ 
\end{tabular}
\end{ruledtabular}
\end{center}
\label{table: YH1 K and w_Wolff L=7}
\end{table*}

\begin{table*}[htb]
%\tiny
\caption{The odd eigenvalue exponent $y_{H1}$ calculated on lattice $256^3$ by using the
Wolff algorithm\cite{Wolff_1989} with different values of $K$ and $u$. $n$ denotes the number of RG iterations.
}
\begin{center}
\begin{ruledtabular}
\begin{tabular}{ccccccccc}
%\hline
$ $ & $K$  & $w$ & $u$ & $n=6$ & $n=5$ & $n=4$ & $n=3$ & $n=2$ \\
\hline						
1 & 0.2216 & 0.5  & 0.8808 & 2.37353(20)  &  2.44672(10) & 2.47271(5) &  2.48299(3) &	2.4974(2) \\
2 & 0.2216 & 0.59 & 0.9137 & 2.37940(19)  &  2.44699(9)	&  2.47025(4) &  2.47696(2) &   2.7979(2) \\		
3 & 0.2216 & 0.7  & 0.9427 & 2.38191(19)  &  2.44766(9)	&  2.46893(4) &  2.47333(2) &	2.47156(1) \\	
4 & 0.2217 & 0.5  & 0.8808 & 2.57931(17)  &  2.51560(6)	&  2.49560(3) &  2.49061(1) &   2.49431(1) \\
5 & 0.2217 & 0.59 & 0.9137 & 2.57317(17)  &  2.51244(7)	&  2.49193(3) &  2.48424(1) &	2.48229(1) \\	
6 & 0.2217 & 0.7  & 0.9427 & 2.56978(17)  &  2.51039(6)	&  2.49003(3) &  2.48036(1) &   2.47398(1)  \\
\hline
7 & 0.221652674 & 0.58870 & 0.91331 & 2.47834(14)  &  2.48115(5) &   2.48166(3) &  2.48087(1) & 2.48127(1) \\
8 & 0.221654712 & 0.58411 & 0.91185 & 2.48225(14)  &  2.48253(6) &   2.48225(3) &  2.48127(1) & 2.48180(1) \\
9 & 0.221654781 & 0.58371 & 0.91172 & & & & & \\
\end{tabular}
\end{ruledtabular}
\end{center}
\label{table: YH1 K and w_Wolff L=8}
\end{table*}

\begin{table}[htb]
%\small
\caption{Estimates of the odd eigenvalue exponent $y_{H1}$ from several sources. 
Values that are boldfaced
are calculated to be consistent with the published exponents in the same source.
}
\begin{center}
\begin{ruledtabular}
\begin{tabular}{cccccc}
%\hline
{\footnotesize{This work}} & Ref.\cite{Ron_Brandt_RHS_2017} & Ref.\cite{Bloete_Heringa_Hoogland_Meyer_Smit_1996} & Ref.\cite{Hasenbusch_2010} & Ref.\cite{Guida_ZinnJustin_1998} & Ref.\cite{Guida_ZinnJustin_1998} \\
MCRG & MCRG & MCRG & MC & $\epsilon$-{\footnotesize{expansion}} & d = 3 \\
\hline

{\footnotesize{2.4824(1)}} & {\footnotesize{2.4829(2)}} & {\footnotesize{2.481(1)}} & {\footnotesize{{\bf{2.4819}}(1)}} & {\footnotesize{{\bf{2.4820}}(25)}} & {\footnotesize{{\bf{2.4833}}(13)}}
 \\

%\hline
\end{tabular}
\end{ruledtabular}
\end{center}
\label{table: YH1 REF}
\end{table}

\section{Summary and future work}
The result of our computation for $y_{H1}$ and a comparison with
other works are shown in Table \ref{table: YH1 REF}. The agreement between the various methods is generally good, although
some differences exist. Since we don’t have estimates of
the systematic errors in our results, we can’t really say
what the source of the differences are.
The most reliable of the estimates shown in Table \ref{table: YH1 REF}
are those of Hasenbusch\cite{Hasenbusch_2010}. This was a very careful
Monte Carlo finite-size study that included many effects
of corrections to scaling to provide limits on the systematic errors.

Our obtained estimate for $K_c=0.2216547(1)$ 
is in agreement with $K_c=0.2216544(3) $\cite{Talapov_Blote_1996} and with 0.22165463(8)\cite{Hasenbusch_2010}. 

The main future work should of course be to extend
the current method to fast calculation of critical properties of other models. 
An essential difficulty is the use of the Wolff algorithm\cite{Wolff_1989}  to generate a set of configurations characterizing the equilibrium distribution, as the Wolff algorithm is not general enough. We would like to develop an algorithm 
which would be 
based on the inverse RG simulations, which only involves simple Monte Carlo as presented in \cite{Ron_RHS_Brandt_2002} and in \cite{Ron_RHS_Brandt_2005}.

\bibliography{MCRGarXiv}% Produces the bibliography via BibTeX.

\providecommand{\noopsort}[1]{}\providecommand{\singleletter}[1]{#1}%
\begin{thebibliography}{10}

\bibitem{Ma_MCRG_1976}
S.-K. Ma.
\newblock Renormalization {G}roup by {M}onte {C}arlo methods.
\newblock {\em Phys. Rev. Lett.}, 37:461--464, 1976.

\bibitem{RHS_MCRG_1979}
R.~H. Swendsen.
\newblock Monte {C}arlo {R}enormalization {G}roup.
\newblock {\em Phys.Rev. Lett.}, 42:859--861, 1979.

\bibitem{RHS_MCRG_optimize_d=2}
R.~H. Swendsen.
\newblock {Monte} {Carlo} calculation of renormalized coupling parameters: {I}.
  d=2 {I}sing model.
\newblock {\em Phys. Rev. B}, 30:3866, 1984.

\bibitem{RHS_MCRG_optimize_d=3}
R.~H. Swendsen.
\newblock {M}onte {C}arlo calculation of renormalized coupling parameters:
  {II}. d=3 {I}sing model.
\newblock {\em Phys. Rev. B}, 30:3875, 1984.

\bibitem{Brandt_Ron_2001}
A.~Brandt and D.~Ron.
\newblock Renormalization multigrid ({RMG}): Statistically optimal
  renormalization group flow and coarse-to-fine {Monte} {C}arlo acceleration.
\newblock {\em J. Stat. Phys.}, 103:231257, 2001.

\bibitem{Ron_RHS_2001}
D.~Ron and R.H. Swendsen.
\newblock Calculation of effective hamiltonians for renormalized or
  non-{H}amiltonian systems.
\newblock {\em Phys. Rev. E}, 63:066128, 2001.

\bibitem{Ron_RHS_2002}
D.~Ron and R.H. Swendsen.
\newblock The importance of multi-spin couplings in renormalized
  {H}amiltonians.
\newblock {\em Phys. Rev. E}, 66:056106, 2002.

\bibitem{Ron_RHS_Brandt_2002}
D.~Ron, R.H. Swendsen, and A.~Brandt.
\newblock Inverse {M}onte {C}arlo renormalization group transformations for
  critical phenomena.
\newblock {\em Phys. Rev. Lett.}, 89:275701, 2002.

\bibitem{Ron_RHS_Brandt_2005}
D.~Ron, R.H. Swendsen, and A.~Brandt.
\newblock Computer simulations at the fixed point using an inverse
  renormalization group transformation.
\newblock {\em Physica A}, 346:387--399, 2005.

\bibitem{Bloete_RHS_3d_Ising_1979}
H.~W.~J. Bl\"{o}te and R.H. Swendsen.
\newblock Critical behavior of the three-dimensional {I}sing model.
\newblock {\em Phys. Rev. B}, 20:2077--2079, 1979.

\bibitem{Pawley_MCRG_3d_Ising_1984}
G.~S. Pawley, R.~H. Swendsen, D.~J. Wallace, and K.~G. Wilson.
\newblock {M}onte {C}arlo renormalization-group calculations of critical
  behavior in the simple-cubic {I}sing model.
\newblock {\em Phys. Rev. B}, 29:4030--4040, 1984.

\bibitem{Bloete_Heringa_Hoogland_Meyer_Smit_1996}
H.~W.~J. Bl\"{o}te, J.~R. Heringa, A.~Hoogland, E.~W. Meyer, and T.~S. Smit.
\newblock {Monte} {Carlo} renormalization of the {3D} {Ising} model:
  Analyticity and convergence.
\newblock {\em Phys.Rev. Lett.}, 76:2613--2616, 1996.

\bibitem{Ron_Brandt_RHS_2017}
D.~Ron, A.~Brandt, and R.H. Swendsen.
\newblock Surprising convergence of the {M}onte {C}arlo renormalization group
  for the d=3 {I}sing model.
\newblock {\em Phys. Rev. E}, 95:053305, 2017.

\bibitem{Baillie_Gupta_Hawick_Pawley_1992}
C.~F. Baillie, R.~Gupta, K.~A. Hawick, and G.~S. Pawley.
\newblock Monte carlo renormalization-group study of the three-dimensional
  ising model.
\newblock {\em Phys. Rev. B}, 45:10438--10453, May 1992.

\bibitem{Wolff_1989}
Ulli Wolff.
\newblock Collective {Monte} {Carlo} updating for spin systems.
\newblock {\em Phys. Rev. Letters}, 62:361, 1989.

\bibitem{Talapov_Blote_1996}
A.~L. Talapov and H.~W.~J. Bl\"{o}te.
\newblock The magnetization of the 3d ising model.
\newblock {\em J. Phys. A:}, 29:5727--5734, 1996.

\bibitem{Hasenbusch_2010}
M.~Hasenbusch.
\newblock Finite size scaling study of lattice models in the three-dimensional
  ising universality class.
\newblock {\em Phys. Rev. B}, 82:174433, 2010.

\bibitem{Guida_ZinnJustin_1998}
R.~Guida and J.~Zinn-Justin.
\newblock Critical exponents of the {N}-vector model.
\newblock {\em J. Phys. A: Math. Gen}, 31:8103--8121, 1998.

\end{thebibliography}

\end{document}